\documentclass[journal]{IEEEtran}

\usepackage{graphicx,amssymb,amsmath}
\usepackage{multicol}
\usepackage[noadjust]{cite}
\usepackage{setspace}
\usepackage{stfloats}
\usepackage{algorithm}

\usepackage{amsthm}
\usepackage{amsmath}
\usepackage{flushend}
\usepackage[table]{xcolor}
\usepackage{cases,subeqnarray}
\usepackage{bm,multirow,bigstrut}
\usepackage{textcomp}
\usepackage{latexsym,bm}
\usepackage{booktabs,changebar}
\usepackage{xcolor}
\usepackage{mathtools}
\usepackage{dsfont}
\usepackage{extarrows}
\usepackage{mathrsfs}
\usepackage{cite}
\usepackage{bm}
\usepackage{cleveref}
\usepackage{algpseudocode}
\usepackage{amsmath}

\theoremstyle{plain}

\theoremstyle{plain}

\IEEEoverridecommandlockouts

\begin{document}

\title{RIS-Aided Next-Generation High-Speed Train Communications: Challenges, Solutions, and Future Directions}
\author {Heng~Liu, Jiayi~Zhang,~\IEEEmembership{Senior Member,~IEEE}, Qingqing~Wu,~\IEEEmembership{Member,~IEEE}, Yu~Jin, Yuanbin Chen, and Bo~Ai,~\IEEEmembership{Senior Member,~IEEE}

\thanks{H. Liu is with the School of Electronics and Information Engineering, Beijing Jiaotong University, Beijing 100044, P. R. China, and also with the State Key Laboratory of Internet of Things for Smart City, University of Macau, Macao, 999078 China. J. Zhang and Y. Jin are with the School of Electronics and Information Engineering, Beijing Jiaotong University, Beijing 100044, P. R. China (e-mail: jiayizhang@bjtu.edu.cn). Q. Wu is with the State Key Laboratory of Internet of Things for Smart City and Department of Electrical and Computer Engineering, University of Macau, Macao, 999078 China. Y. Chen is with the State Key Laboratory of Networking and Switching Technology, Beijing University of Posts and Telecommunications, Beijing 100876, China. B. Ai is with State Key Laboratory of Rail Traffic Control and Safety, Beijing Jiaotong University, Beijing 100044, China.}

}
\maketitle

\begin{abstract}
High-speed train (HST) communication is a very challenging scenario in wireless communication systems due to its fast time-varying channels and significant penetration loss.
Thanks to the low cost and smart channel reconfiguration ability, reconfigurable intelligent surface (RIS) has been proposed as a promising technology to solve the aforementioned challenges cost-effectively and to achieve high spectral and energy-efficiency in the next-generation HST communication systems.
In this article, we first provide a novel RIS-aided HST wireless communication paradigm, including its main challenges and application scenarios.
Then, we propose practical solutions are provided to realize efficient signal processing and resource management in the considered system.
Finally, future directions of RIS-aided HST communication systems are pointed out to inspire further investigation in future work.
\end{abstract}

\IEEEpeerreviewmaketitle

\section{Introduction}
Recently, wireless communication in high-speed train (HST) systems has evolved from traditional low-data rate control signaling to diversified data-intensive services, such as high-definition video surveillance, onboard broadband internet services, and the railway Internet of Things (IoT) service \cite{ai20205g}.
To guarantee safe and reliable train operations, and support broadband internet access for high-speed passengers, emerging applications on HST requiring seamless coverage are unavailable for high mobility. Unlike other wireless networks, HST wireless systems are characterized by the high mobility of onboard transceivers and the large signal penetration loss passing through train carriages. These special characteristics bring many design challenges such as channel modeling, doppler shift compensation, time-varying channel estimation, beamforming design, and resource management.

To well support the demand of large data traffic and bandwidth-intensive applications and solve the challenges in HST communications, massive multiple-input multiple-output (MIMO) and millimeter wave (mmWave) communication are deemed as technical support for on-going fifth- generation (5G) cellular system for railway. However, exploiting a large number of active antennas as in massive MIMO and mmWave requires considerably high hardware cost, energy consumption, as well as computational complexity. Furthermore, in mmWave communication systems, the propagation environment is uncontrollable. Therefore, the quality of service (QoS) is significantly degraded when the line-of-sight (LoS) mmWave links between BS and user equipments (UEs) are blocked by the HST carriages. Finally, it is also difficult to effectively mitigate the large doppler effect using existing technologies.

Considering the aforementioned issues and challenges, the recently proposed reconfigurable intelligent surface (RIS) emerges as a promising technology for HST communication systems to achieve high spectral- and energy-efficiency. It can also extend the coverage with low cost, complexity and energy consumption \cite{di2020smart}, \cite{huang2020holographic}.
In RIS-aided HST systems, BS can be equipped with a small number of antennas to achieve the required QoS, to reduce the energy consumption and hardware cost.
Moreover, RIS can be properly deployed to combat the blockage of mmWave communications by creating additional virtual LoS path \cite{zhang2020prospective}.
The reconfigurability of RIS on the amplitude and phase shift can also be utilized to solve the doppler effect and multipath fading problems.
In addition, the beamforming of RIS can increase signal gain and reduce the influence of signal attenuation caused by train carriages.
It is also appealing to limit the signal-leakage outside the intended coverage area and extend the system coverage to decrease the frequent handover.
The maturity of the HST IoT industry chain has ramified a variety of smart rail services and applications. For instance, the train operation control system and real-time communication between the train and trackside equipment, have extreme requirements for reliability and latency. Passenger amusement services require a high rate of Internet access. Thanks to the reflected/refracted high beamforming gain of the RIS, we expect more excellent performance in the RIS-aided HST communication system in terms of QoS indicators (such as high reliability, low latency, and ultra-large capacity connections).

To fully utilize the RISs' benefits and solve the challenging HST communication problem, this article for the first time investigates RISs for the HST communication systems, highlighting the key challenges and a range of promising future research directions.
Specifically, in the following, we first investigate the main application scenarios of RIS-aided HST communication systems to decrease the effect of doppler and penetration loss and improve the performance gain.
We study how we can exploit the machine learning technologies to design the efficient channel estimation, beamforming optimization, and resource management algorithms by considering the characteristics of the HST system and provide the numerical results to validate the various benefits of RIS brought for HST communication systems.
Finally, we also highlight research directions worthy of further investigation.
\section{Challenges of RIS Aided HST Communication Systems}\label{HST:challenges}
\subsection{Challenges of HST communication systems}
The demand for communication is essential for the implementation of intelligent rail. The highly reliable and low latency is very important for the control signal to maintain the safety and reliability of the train operation, even to realize the automatic train operation. Furthermore, high rates communication links are required to support data-demanding applications, such as high definition video to keep the train monitor and improve the passengers' comfortable Internet experience. Besides, the cost of intelligent train is expected to be lower for large energy efficiency and environmental friendliness.

HST communication is a challenging scenario in the communication system as it has several unique characteristics.
The large doppler effect caused by high-speed movement is always an inevitable problem that needs to be carefully handled.
Besides, the signal attenuation caused by the train carriage will degrade the system performance.
Moreover, the fast time-varying and nonstationary channels are difficult to be accurately estimated.
More importantly, beamforming technique strongly depends on the channel state information (CSI) which could become outdated quickly in high mobility scenarios, because the corresponding coherence time is much shorter in the rapidly time-varying channel.
Finally, the frequent handover is another challenge for multi-users to be accessed to a new BS simultaneously.

Since the increased users and the diversified services in HST render time-frequency resources to become scarcer, an effective strategy of resource sharing is essential to moderate the congestion of services. In terms of the physical layer, although the aid of RIS has enhanced the directionality of the signal, as several transceivers aided by RISs share the spectrum resources, the multiple access method should be clarified. In addition, RIS can alleviate the interference resulted from resource reusing by properly adjusting the reflection coefficients. When a certain frequency band is occupied, reasonable resource management can ensure each HST service's continuity and QoS.

On the other hand, the special high-speed railway scenarios also bring new design opportunities.
specifically, the characteristics of determined running directions and regular running tracks lead to a regular and predictable doppler shift curve, thereby making it easy to track and compensate the doppler effects.

\subsection{Challenges of RIS Implement}

\begin{figure}[t!]
\centering
\includegraphics[scale=0.6]{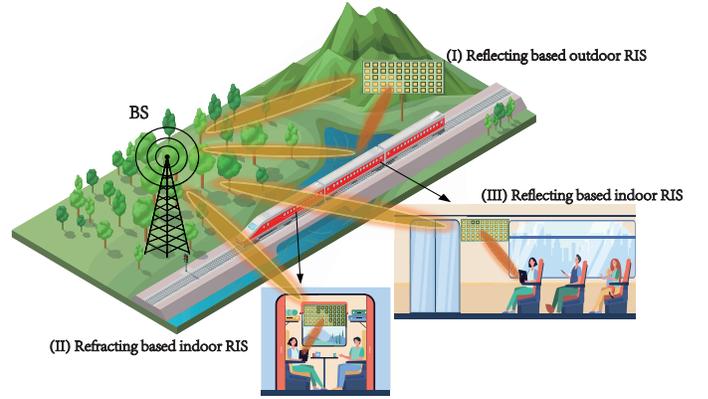}
\caption{Main application scenarios for RIS-aided HST communication systems. (I) Reflecting surface deployed at the trackside, (II) Refracting surface attached to the windows of the train, and (III) Reflecting surface deployed on the wall inside the train.
\label{fig:model}}
\end{figure}
RIS can be either a passive reflect or transmissive planar array made of two-dimensional metamaterial which can be digitally controlled. The reconfigurable complex coefficient of the $n$-th element of the RIS is denoted as ${\beta _n}{e^{j{\phi _n}}}$, where ${\beta _n}$ and $\phi_n $ are the reconfigurable amplitude and phase shift, respectively.
It is worth pointing out that RIS here refers to not only the reflection-based metasurface but also the refraction-based one, which can be made transparent and attached to windows of the train.
Taking the practical hardware architecture of RIS into account, the discrete amplitude and phase shift, elements' mutual coupling, and hardware impairments should be considered in the practical implement \cite{wu2019towards}.

RISs are very appropriate for HST communication systems, since it can provide virtual LoS path with less energy consumption to enhance the effective channel power gain. Furthermore, it can provide solution to solve the doppler effect caused by the mobility of the train. Moreover, it is easy to deploy in the current HST system without too much change in existing communication systems. Although there has been upsurge of interest in applying RIS in wireless communications, the study of RIS-assisted HST communications is still in a very early stage and many interesting and important problems need to be addressed.
\section{Main Application Scenarios}
RIS can be flexibly deployed in the current communication systems. In traditional low-mobility scenarios, RIS is often deployed nearby the BS or UE to have the LoS path, which can extend the coverage area and provide wireless services for deadzone areas when the direct link is blocked by an obstacle. In the HST systems, considering the practical factors such as deployment/operational cost, user demand/distribution, space constraint, as well as the propagation environment, three different application scenarios can be considered in the RIS-assisted HST communication system.

One straightforward scenario is to deploy RIS on one side of the railway, for example, on a roadside advertising board, as a reflective panel to reflect signals. The location of RIS is deployed between the two BSs. This deployment can expand the coverage of a single cell to reduce frequent cell handovers. In addition, due to the participation of RIS in the system, the deployment distance between BSs can be increased, thereby reducing the hardware costs caused by the use of active BSs. In addition, since the reflection path can provide a strong LoS path, it can also solve some blockage problems. This method can be easily deployed and will not have a significant impact on the existing HST communication system. Finally, the frequent handover problem (e.g., hundreds of UEs need to be handed over to the next BS in a very short time) can be eliminated in a wide area. Although the trackside-based RIS is a robust and flexible architecture for supporting HST communications in a small network, it is generally unable to provide a scalable solution for serving UEs due to the movement of the train. Despite the promising advantages of trackside RIS-aided HST communication systems, there are still some train operational environments where the trackside RIS are unavailable, such as in tunnels.

From another perspective, RIS can be attached to the windows of the train to transmit the signal incident on the window. In this scenario, the RIS is a transparent surface, which can control the refractive coefficient to provide better service. The path-loss between RIS and UE is small as the distance between them is small, which brings the advantage of decreasing the signal attenuation. Since the fixed paths between the RIS and the UEs in the train make channels quasi-static, the large time scale beamforming can be utilized to provide the performance gain, which is a practical method to adjust the refractive coefficients with low complexity. The control signal of the RIS will be received by the train. However, the refraction amplitude may be significantly smaller than that of reflection based due to the penetration loss passing through the windows. Moreover, the size of RIS is generally limited by the size of windows and the LoS path between the BS and the window-based RIS may be blocked by the obstacles.

Similar to the conventional indoor scenario, RIS also can be attached to the wall in the train to help enhance the received signal at UE. The signal transmitted from the BS will reach the RIS through the window, and then RIS will configure the reflect coefficients and reflect the signal to receivers. Therefore, this application scenario is suitable for any train operational environment. However, the signal reached the RIS is weak due to signal attenuation caused by the train body and windows.
Finally, it is envisioned that the future wireless network for supporting RIS-aided HST communication systems will have an integrated 3-D architecture consisting of trackside-based RIS, window-based RIS, and wall-based RIS, as shown in Fig. \ref{fig:model}.

\begin{figure}[t]
\centering
\includegraphics[scale=0.55]{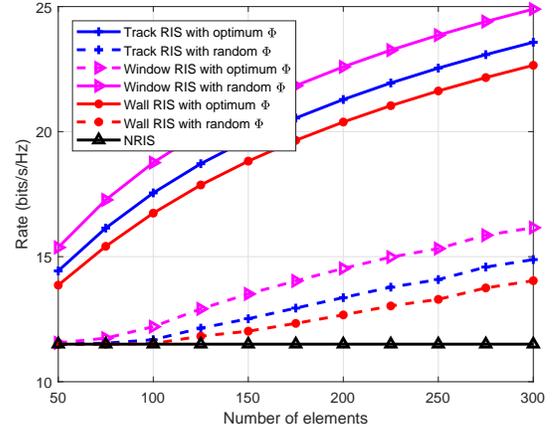}
\caption{Effect of the number of RIS elements for different scenarios.
\label{fig:ELEMENT}}
\end{figure}
We consider a 3D model with a BS with a single antenna, an RIS with $N$ elements and one single-antenna user. We assume that the coordinate of BS, the trackside-based RIS, the window-based RIS, the wall-based RIS, and UE is (0,0,25), (0,60,25), (9,55,0), (11,55,0) and (10,50,0), respectively. It is assumed that all the channel following the Rician fading channel with path loss exponent of 2.2. The receiver noise power is -80 dBm. The transmit power is 10 dBm. The number of RIS elements is 100. We set the amplitude of trackside-based RIS to 1, the amplitude of window-based RIS and wall-based RIS to 0.25 due to the signal penetration loss of the surface and window.

Fig. \ref{fig:ELEMENT} illustrates the influence of different numbers of RIS elements on the system performance of different scenarios. Compared to the scenario with no-RIS (NRIS), the performance of all scenarios with RIS improved as the
numbers of RIS elements is increased. Also, the performance gaps among these scenarios increase as the number of elements grows. Moreover, even if the number of the RIS elements is very small with 50 elements, our proposed
scenario of wall-based indoor RIS with optimum $\mathbf{\Phi}$ still have at least 19$\%$ performance gain which shows the effectiveness of the proposed scenarios. Furthermore, the proposed window-based RIS scenario outperforms other scenarios since the RIS-UE channel experiences small path loss, which indicates that it is better to deploy RIS on the window in the practical rural case.

\section{Promising Solutions}
In this section we consider the practical RIS-aided HST communication system from a signal processing perspective, highlighting the key challenges and potential solutions. In the following, we will discuss the issues of channel estimation, beamforming design, and resource management, respectively. Table \ref{tab:tab1} summarizes the promising solutions of RIS-aided HST communication systems, which are further elaborated as follows.

\begin{table*}[t]
\newcommand{\tabincell}[2]{\begin{tabular}{@{}#1@{}}#2\end{tabular}}  
\centering
\caption{Comparison of different Signal Processing algorithms}\label{tab:tab1}
\begin{tabular}{|c|c|c|c|c|}\hline
\rowcolor{gray!20} \textbf{Methods} & \textbf{Schemes} & \textbf{Algorithms} & \textbf{Characteristics} & \textbf{Practicality of HST}\\\hline
\multirow{5}{*}{\tabincell{c}{Channel \\ Estimation}} & \multirow{3}{*}{ \tabincell{c}{Passive, \\ Semi-Passive}}      & MMSE           & \tabincell{c}{Large complexity, high \\ performance and high overhead}  & Not practical \\\cline{3-5}
                                    &                               & Compress sensing \cite{taha2019enabling} & \tabincell{c}{Moderate complexity, better \\ performance and moderate overhead}  & Practical\\\cline{3-5}
                                    &                               & Deep learning \cite{taha2019enabling}    & \tabincell{c}{Moderate complexity, better \\performance and small overhead}   & Practical\\\hline
\multirow{4}{*}{Beamforming} & Perfect CSI      & AO \cite{wu2019intelligent}  & Low complexity& Not practical    \\\cline{2-5}
                                    & \multirow{3}{*}{Robust}       & Two-time scale \cite{zhao2020intelligent}  & Moderate complexity & Practical    \\\cline{3-5}
                                    &                               & CSI error based AO \cite{zhou2020framework} & High complexity & Practical    \\\cline{3-5}
                                    &                               & Deep learning \cite{huang2020reconfigurable}                    & High complexity & Practical    \\\hline
\end{tabular}
\end{table*}

\subsection{Channel Estimation}
The acquisition of accurate CSI is very important to achieve the promising superior performance gains. However, it is challenging to obtain the accurate CSI because the additional reflect paths brought by RIS and the fast varying channels caused by mobility bring difficulty to obtain the accurate CSI in RIS aided HST communication systems, especially when the number of elements is practically large.
Traditionally, the BS-RIS-UE channel can be estimated as a cascaded channel by sending the pilot sequences and processing the signal at the receiver side. Moreover, it can also be estimated separately by adding some RF chains on the semi-passive RIS. However, this will be at the cost of higher energy consumption and computational complexity.

Therefore, combined with the features of the HST communication, some prior information can be utilized to decrease the estimation complexity and reduce the training overhead. The past location, the direction and the velocity can be used to predict the following location of the train. The information on the channel parameters, e.g., angles of departure and angles of arrival, can be derived from an external location information system on the BS and the train.

In different scenarios, different special channel characteristics can be utilized to design the channel estimation protocol to reduce pilot overhead. In the track-side RIS scenario, since the positions of the BS and the RIS are fixed, the channel between the RIS and the BS generally remain quasi-static, while the channel between the RIS and the UE and the channel between the BS and the UE changes rapidly over time due to the mobility of the UEs. Therefore, high-dimensional quasi-static BS-RIS can be estimated in a larger time scale, and low-dimensional mobile RIS-UE and BS-UE channels can be estimated in a smaller time scale. This two-time scale channel estimation framework can significantly reduce pilot overhead \cite{wei2021channel}. Correspondingly, in the window RIS and wall RIS scenarios, the two-time scale channel estimation framework can also be used. Furthermore, as the UEs are located in a cluster, the anchor node can be deployed at the center of the train to estimate the common channels, which can be used to reduce the real-time training overhead for estimating the other UEs' channels.

The traditional least square (LS) and minimum mean square error (MMSE) channel estimation methods can be used to estimate the channel with huge computational complexity. Then the compress sensing based channel estimation can be used by leveraging the sparsity property. Besides, the channel matrix can also be viewed as a noise-based figure, which can be denoised by utilizing the deep learning approach. Such scenarios motivate the use of deep learning to offline estimate the channel based on the historical coherent channel statement and the location information just as the video prediction. The idea of the convolutional neural network method is to map iterative algorithms to deep neural networks. It is noted that hybrid neural networks and nonlinear continuous output can be used to improve the estimation accuracy and adapt to high-speed moving time-varying channels. Moreover, the use of continuous nonlinear joint loss function can be used to repair the influence from doppler shift.

\begin{figure}[t]
\centering
\includegraphics[scale=0.55]{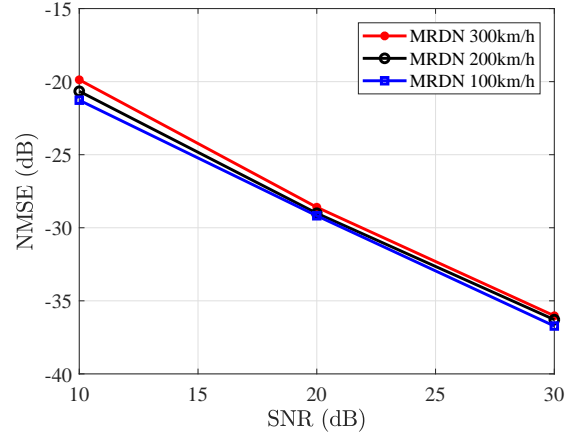}
\caption{NMSE versus SNR with different train speed.
\label{fig:nmse}}
\end{figure}

Fig. \ref{fig:nmse} illustrates the normalized mean squared error (NMSE) performance of the proposed multiple residual dense networks (MRDN) based channel estimation with different train speeds. Note that the MRDN exploits the low channel sparsity characteristic to recover the HR channels. It can be observed that the NMSE performance grows with the increasing signal-to-noise ratio (SNR). For different operation speeds, the channel estimation algorithm can bring a similar performance, which indicates the efficiency of the proposed channel estimation algorithm.

\subsection{Beamforming Design}
With the aid of estimated CSI that will be encountered during the transmission phase, a substantial performance improvement can be provided for RIS-aided HST communication systems by employing beamforming design. How to well design reconfigurable amplitudes and phase shifts of the RIS to enhance the received signals and reduce the interference is important but difficult. Second, power control is another useful scheme to mitigate interference. The beamforming needs to be jointly optimized with the transmit beamforming of the active BS and the passive RIS to improve the system performance.

The existing beamforming designs assumed that the perfect CSI can be obtained in low mobility scenarios, which can not be used in the high mobility scenario. In high mobility scenarios with fast time-varying channels, tuning RIS parameters via exploiting instantaneous CSI is challenging due to the following reasons. First, in a short duration of coherence time, system resources have to be reallocated and the RIS phase shift parameters have to be updated frequently, thus incurring significant signaling overhead.
Second, the instantaneous CSI is very difficult to obtain. Besides, the doppler effect also needs to be considered in the beamforming to decrease the doppler effect and improve the system performance. Rapid fluctuations in the received signal strength due to the doppler effect can be effectively reduced by using the real-time tuneable RISs \cite{basar2019wireless}.

Hence, the robust beamforming design can be used because it can bring the performance gain within tolerating a certain degree of channel estimation error. As for the robust beamforming design, two kinds of CSI error models can be considered, namely the bounded CSI error model and statistical CSI error model \cite{zhou2020framework}. And there are different kinds of optimization objects. In the error limited scheme, the object is to maximize the minimum SINR to guarantee the QoS requirements. Another is to guarantee the outage probability. Besides, the two-time scale beamforming design which considers long-term RIS phase shifts optimization based on statistical CSI and the short-term transmit
precoding vectors design at the BS with the effective CSI, is also suitable in the refractive scenario \cite{zhao2020intelligent}. By exploiting useful information such as location and mobility pattern, the beam can be designed to track the movement of the train, which helps reduce the complexity for beamforming \cite{matthiesen2020intelligent}.

Due to the multi-valuables coupled in the optimization object, the alternative optimization (AO) can also be used in such scenarios to decouple the valuables. Then the optimized variables can be optimized in turn until the optimization result is converged. Besides, deep reinforcement learning (DRL) can be leveraged to learn and build knowledge about the varying channels and the environment. Then, efficient algorithm can be designed by observing the rewards from the environment and proper actions can be chosen to find out the solution of the optimization problems \cite{huang2020reconfigurable}. It is used to obtain the beamforming matrix by mapping relations between the channel environment and the precoding design.

\begin{figure}[t]
\centering
\includegraphics[scale=0.55]{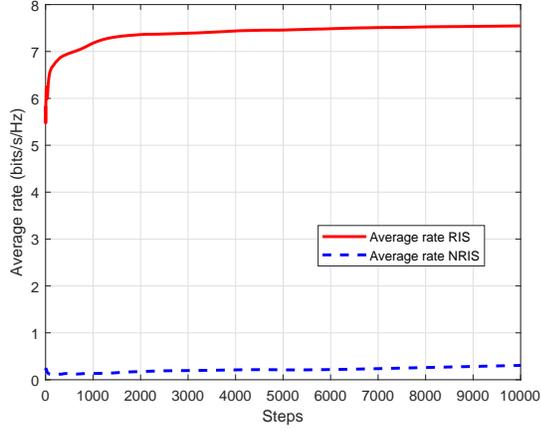}
\caption{Rate versus training steps for different scenarios.
\label{fig:bf}}
\end{figure}

In Fig. \ref{fig:bf}, the rate performance of the optimization is illustrated with different settings. Specifically, DRL approach is utilized to solve the sum rate maximization problem by considering the channel and location information as states and the beamforming values as actions. It can be observed that the solved results can be obtained with several iterative steps by utilizing the DRL algorithm. The performance of RIS assisted scenario outperforms the NRIS assisted scenario, which highlights the necessary of investigating RIS into the HST communication systems.

\subsection{Network Slicing Enabled Resource Management}
We have discussed some high mobility-induced issues in the RIS-aided HST communication systems above, such as Doppler frequency, penetration loss, and frequency handover. Another pivotal challenge in terms of the optimization of the HST communication network is to delicately manage the time-frequency resources, which can greatly guarantee the QoS requirements of each HST service and achieve a high spectrum resource utilization.

Different priorities of the HST services imply unfair competition and resource utilization among the users. To circumvent this issue, it is suggested to integrate the cognitive radio (CR) technology in the RIS-aided HST system. In particular, CR refers to the sharing of spectrum resources between unlicensed users (also known as secondary users, SUs) and licensed users (also known as primary users, PUs), where PUs own higher priority. SUs are required to discover holes in time-frequency resources and BSs are required to recognize time-frequency resource environments when accessing the spectrum. Thus, we regard BSs with cognitive and computing capabilities as time-frequency resource management equipment. Since the train operation control system has the highest requirements for safety, fixed spectrum resources should be allocated, that is, neither PU nor SU are they. Due to the spectrum sharing between PUs and SUs, the QoS of PUs is inevitably weakened by the interference from SU. The deployment of RIS can effectively suppress interference and guarantee the QoS performance of the total system.

A logically isolated and customized dedicated network is necessary for each HST service to satisfy its QoS requirements. In this regard, network slicing technology has come forward. Different reusing rules of time-frequency resources in CR can inspire us to blend different virtual resources and physical facilities into different slicing. The divided slicing consists of the resource sets at the radio access network (RAN) and the core network (CN), such as virtual time-frequency resources, RIS elements, fronthaul/backhaul links, core network servers, etc., which requires a unified mapping relationship to reflect it to an end-to-end logical network. Furthermore, we can design the Internet access slicing and IoT application slicing, which can be attributed to the category of PU, and design the roadside equipment service slicing that belongs to the category of SU. Particularly, the train operation control service that has the highest security level, holds its dedicated slicing. The above-mentioned slicing with physical facilities and virtual radio resources furnishes isolated, secure, and highly customized end-to-end logical networks for different types of service demanders, such as vertical industry users, virtual operators, and enterprise terminals.

Since the discussion of network slicing technology in the RIS-aided HST communication systems is still in its infancy, there are still some challenges as follows:
i) Dynamic slicing orchestration and management: A major challenge for network slicing to enable RISs-aided HST communication systems is to map from the service application layer to the specific physical infrastructure. In addition to mapping RBs in RAN and routing severs in CN into different slicing, when the HST is equipped with a large RIS, the tiles \cite{najafi2020physics} composed of different groups of RIS elements may serve different users, so different tiles can be accommodated in certain slicing. It is surely a more straightforward idea to map different RISs into specific slicing.
ii) Mobility management technology: The ultra-high mobility of HST leads to the challenge of frequency handover of BSs and RISs. Fast mobility handover is the key to ensuring the QoS of services. The end-to-end network slicing can cover not only the resources of BS and RISs but the resources within the service range of multiple infrastructures. Hence, reasonable granularity and migration management scheme of the network slicing can avoid high overhead and ensure seamless connection to fully satisfy QoS requirements.
iii) Appropriate resource allocation algorithm: Effective interference management is necessary since different reflected/refracted beams of RISs are directed to different user terminals. Time, frequency, power, beam, and other resources can be weighed to design a multi-dimensional domain adaptive resource scheduling algorithm for intra-/inter slicing to meet the QoS requirements of multiple services at the same time.

\begin{figure}[t]
\centering
\includegraphics[scale=0.55]{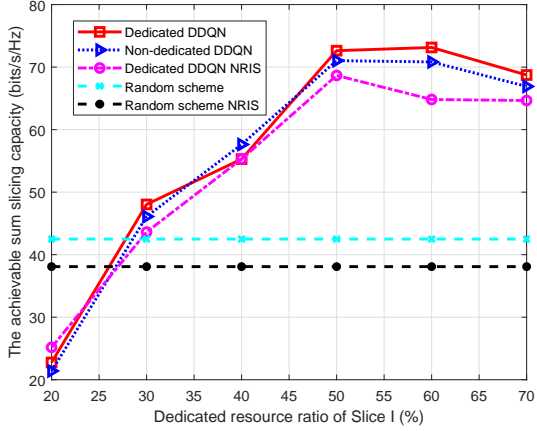}
\caption{The achievable sum slicing capacity versus the recourse ratio of slicing I for different scenarios.
\label{fig:RRM}}
\end{figure}

Three types of typical network slicing are considered in the simulation, which are the train operation control slicing (Slicing I), the amusement slicing (Slicing II), and the slicing that carries services with higher levels of safety and reliability (Slicing III), respectively. Since the train operation control system has the highest priority, the network resources (including resource block, computing servers and RISs) allocated to Slicing I are not reused by any other slicing while the remaining idle resources are shared by the other two slicing. Fig. \ref{fig:RRM} plots the achievable sum slicing capacity versus the dedicated resource ratio of Slicing I. Particularly, the DRL approach is leveraged to enable the resource management for the RIS-aided HST communication systems. As a comparison, several special schemes are proposed. Dedicated Double Deep Q Network (Dedicated DDQN): DDQN framework is adopted and the dedicated resources are allocated to Slicing I; Non-dedicated DDQN: the available network resources are shared by all slicing according to DDQN; Random scheme $\&$ Random scheme NRIS: available network resources are randomly allocated in the case of with/without RIS. It can be seen from Fig. \ref{fig:RRM} that the system revenue of the dedicated slicing scheme is better than that of the non-dedicated scheme. The achievable sum slicing capacity increases first and then decreases, reaching a peak at a $50\%$ dedicated ratio. This can be attributed to the fact that when Slicing I occupies an excess of resources, the available resources acquired by other slicing are squeezed, which falls off the system revenue. The dedicated resource ratio should be traded off to balance the system revenue. In addition, the random schemes are not affected by the dedicated ratio owing to the fair competition among the available network resources.

\section{Future Directions}
In addition to the challenges and solutions discussed above, there are some other research directions worthy of future investigation for the RIS-aided HST communication systems.
\subsection{Channel Model}
The practical channel model should be properly established. Accurate channel modeling is crucial for performance analysis \cite{tang2020wireless}. The far-field channel model is used in most of the existing channel
models to reveal the propagation characteristics in RIS-aided wireless systems. In practice, as we discussed above in order to achieve the best performance, the RIS may be deployed in the vicinity of BSs and/or users. Additionally, RIS may cover a large area to capture a large number of incident electromagnetic waves. In this case, the distances from the BS to all the reflecting elements are different, and a near-field channel model should be adopted. Due to the complex and diverse railway operation environment and the introduction of additional channels brought by RIS, how to establish an accurate channel model in such a rapidly changing environment should be highlighted.
Therefore, it is necessary to build new channel measurement campaigns and modeling approaches to characterize railway channels.
\subsection{Deployment and Control}
Although we have proposed several application scenarios, there are still many problems in practical deployment and control. Due to different railway operating environments, different types of RIS deployment methods must be designed for specific environments to ensure that specific QoS requirements are achieved. In these different deployment methods, how should the control protocol be designed to deal with the RIS deployment in different scenarios. How to achieve synchronization between different devices to ensure the performance of the system. In traditional scenarios, RIS is usually deployed centrally. In high-speed scenarios, when the size of RIS is limited, distributed deployment may bring potential advantages.

Due to the movement of the train, the transmission of control signals and the design of the control protocol are also critical. How to integrate the RIS into the HST systems via protocol design is also critical.
How to coordinate RIS with BSs also needs further study. For some indoor scenarios, wired connection can be used, whereas for outdoor cases, wireless backhaul is practically appealing.
Moreover, the secure transmission protocol design for ensuring the physical layer security is desired in future work.

\subsection{Hardware imperfections}
The conventional system models for RIS-aided communications have been developed based on the assumption of perfect hardware. However, both RIS and its controllers are preferably fabricated using low-cost components which will be subject to hardware impairments.
The hardware impairments will degrade the system performance. The challenges and potential topics with non-ideal RIS signal models should be highlighted. The actual hardware impairments need to be considered, including the limited resolution and the mutual coupling of the reconfigurable amplitude and phase shifts with respect to incident angles. Besides, the backhaul limitation of the control signal to make the real-time RIS configuration requires dedicated research efforts to solve it for further investigation.

\section{Conclusions}\label{se:conclusion}
This article studies the challenges of the RIS-aided HST communication systems and proposes three novel application scenarios to improve the system performance. The RIS is utilized to decrease the doppler effect and reduce the signal attenuation to improve the performance gain in the HST communication systems. Advanced signal processing and recourse management techniques are leveraged to boost the system performance. This article serves as a humble attempt to provide useful directions and insightful inspiration to future research on RIS-aided HST communication systems.
\bibliographystyle{IEEEtran}
\bibliography{IEEEabrv,Ref}

\end{document}